\documentclass[aps,prl,twocolumn,superscriptaddress,nopacs]{revtex4}
\usepackage[pdftex]{graphicx}
\usepackage{amssymb,amsfonts,amsmath}
\usepackage{float}

\newcommand{\cd}{\emph{C.~Diff. }}

\begin{document}

\title{Spread of pathogens in the patient transfer network of US hospitals}

\author{Juan Fern\'andez-Gracia$^{\oplus}$}
\affiliation{Department of Epidemiology, Harvard T.H. Chan School of Public Health}
\affiliation{Instituto de F\'isica Interdisciplinar y Sistemas Complejos IFISC (CSIC-UIB)}
\author{Jukka-Pekka Onnela$^{\oplus}$}
\affiliation{Department of Biostatistics, Harvard T.H. Chan School of Public Health}
\author{Michael Barnett}
\affiliation{Brigham and Women's Hospital}
\author{V\'ictor M. Egu\'iluz}
\affiliation{Instituto de F\'isica Interdisciplinar y Sistemas Complejos IFISC (CSIC-UIB)}
\author{Nicholas A. Christakis}
\affiliation{Yale Institute for Network Science, Yale University}

\noindent \small{$^{\oplus}$ JFG and JPO are joint first authors of this article.}
\vspace{3mm}

\begin{abstract}
Emergent antibiotic-resistant bacterial infections are an increasingly significant source of morbidity and mortality. Antibiotic-resistant organisms have a natural reservoir in hospitals, and recent estimates suggest that almost 2 million people develop hospital-acquired infections each year in the US alone. We investigate a network induced by the transfer of Medicare patients across US hospitals over a 2-year period to learn about the possible role of hospital-to-hospital transfers of patients in the spread of infections. We analyze temporal, geographical, and topological properties of the transfer network and demonstrate, using C.~Diff.~as a case study, that this network may serve as a substrate for the spread of infections. Finally, we study different strategies for the early detection of incipient epidemics, finding that using approximately 2\% of hospitals as sensors, chosen based on their network in-degree, results in optimal performance for this early warning system, enabling the early detection of 80\% of the C.~Diff.~cases.
\end{abstract}


\maketitle

Every year in the US alone, there are 1.7 million nosocomial infections and 99,000 associated deaths, imposing substantial clinical and financial costs to the US health care system \cite{cite1,cite2,cite3}. The vast majority of these are due to antibiotic-resistant bacteria \cite{cite4}, which have a natural reservoir in hospitals, presenting a potentially lethal threat to already-sick patients. The annual cost of antibiotic-resistant infections in the US has been estimated to range from \$21 billion to \$34 billion \cite{cite5,cite6,cite7}. A 2013 CDC (Centers for Disease Control and Prevention) report on antibiotic-resistant bacteria identified the lack of infrastructure to detect and respond to emerging resistant infections as a pressing gap.

Antibiotic-resistant organisms have a natural reservoir in hospitals. In our study, over a two-year period, there were nearly one million transfer events across US hospitals of Medicare patients alone. Given this large number of transfers, the network of patient transfers could plausibly act as a conduit for antibiotic-resistant bacteria from hospital to hospital. There are, however, only a few existing studies that have investigated the possible role of hospital-to-hospital transfers of patients for the spread of infections. Some studies have focused on the structure of the nationwide transfer network associated with critical care  \cite{cite8,cite9,cite10,cite11}, while others have had a more restricted scope, limited to smaller geographical units, such as counties \cite{cite12,cite14}.

Local containment of antibiotic-resistant bacteria at the level of individual hospitals is a difficult but manageable task given that interactions between hospital wards are relatively structured and confined spatially \cite{cite15,cite16}. But controlling a larger epidemic of antibiotic-resistant bacteria or responding to new mass outbreaks is much more challenging. This is in part related to the complex pattern of patient movements between hospitals, which gives rise to a broad, distributed network. To better understand the role of patient transfers for the spread of infections, we pursue three interconnected aims. First, we investigate the structure of the hospital-to-hospital patient transfer network in the US; second, we correlate the incidence of nosocomial infections on a national scale with properties of this network; and third, we develop a scalable method for the efficient early detection of the spread of nosocomial infections.

We start with structural analyses by first aggregating patient transfers over time to create hospital-to-hospital connections (``edges'') in the network, and we then examine static structural properties of this network. We then demonstrate that the transfer network is a plausible substrate for pathogen spread by analyzing the test case of the common and highly transmissible health-care associated infection \emph{Clostridium difficile} (\cd), and showing that \cd incidence in a sample of 21 million hospital visits across the US is correlated with the topology of the patient transfer network. Finally, we propose a system of using a subset of the hospitals as networkâ ``sensors'' that might be used to monitor the nationwide hospital system.

\section{RESULTS}

\subsection{Properties of the transfer network}
The transfer network shows strong seasonal, monthly, and weekly cycles of patient transfers. The topology of the network and the geography of patient transfers are closely related, with 90\% of transfers between hospitals less than 200km apart. On average, over the 2-year period, a hospital sent patients to 13.55 $\pm$ 0.15 (SE) hospitals and received patients from 13.55 $\pm$ 0.25 hospitals. (Note that the two means necessarily coincide in a directed network because each edge has both an outgoing end and an incoming end.) The average number of patients transferred per edge in the 2-year period was 12.3 $\pm$ 0.63 (SE). Although the degree distributions (in-degree and out-degree) have fat tails (more so the in-degree), comparisons of the average clustering coefficient and the average shortest path length to randomized versions of the network show that the network closely resembles a spatial network. 
In particular, it is much more clustered than a random network and has a high average shortest path length. Finally, the network shows no significant assortativity by degree. A representation of the aggregated network is shown in Fig.~\ref{Fig1}. (See the appendices for more details.)

\begin{figure}
\begin{center}
 \includegraphics[width=8.6cm]{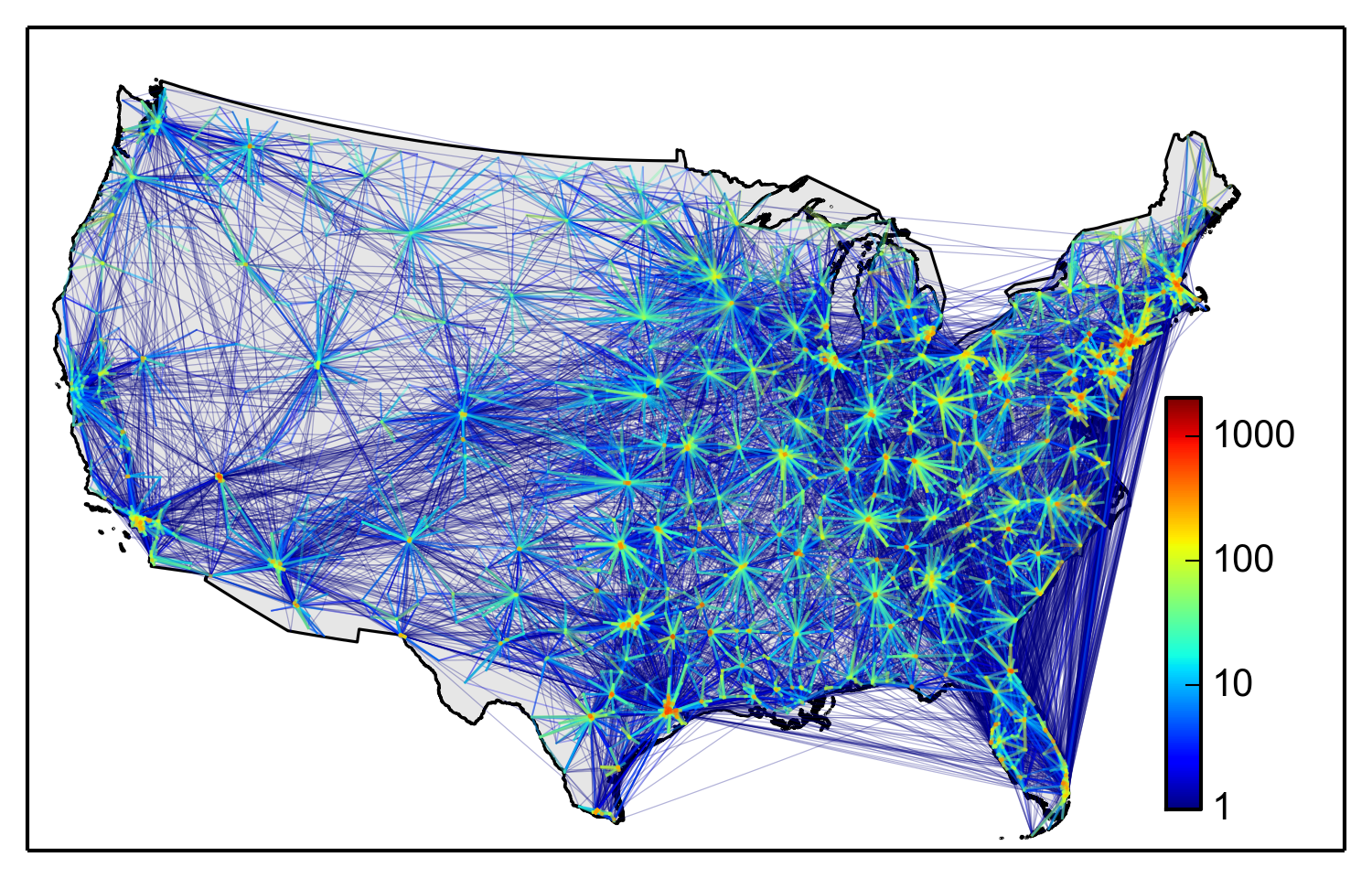}
 \caption{\textbf{Network of hospital-to-hospital transfers of US Medicare patients.} The network consists of hospitals that are connected by daily transfers of patients, here aggregated over the two-year period. Edge color encodes the number of patients transferred through each connection.\label{Fig1}}
 \end{center}
\end{figure}

\subsection{Spread of C.~Diff.~infections}
\begin{figure}
\begin{center}
 \includegraphics[width=8.6cm]{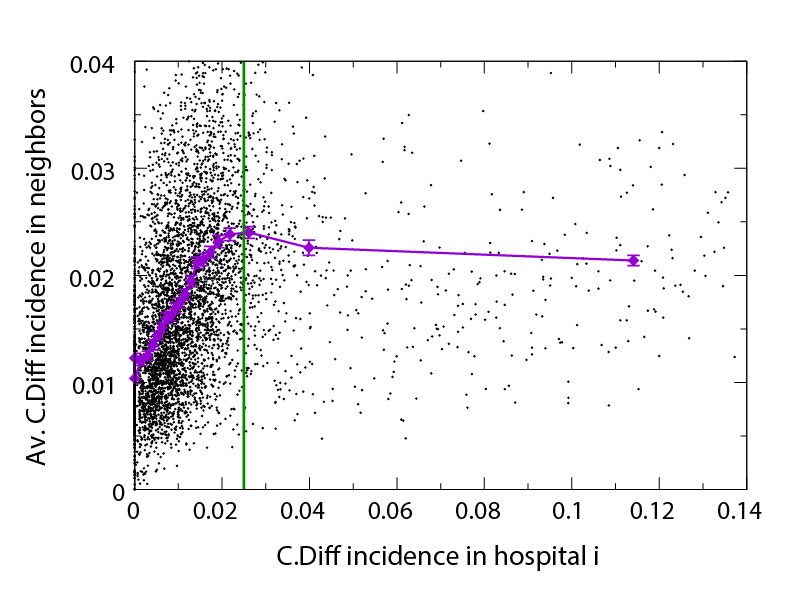}
 \caption{\textbf{Correlation between C.diff. incidence and transfer network structure.} The horizontal axis represents the mean \cd incidence at the focal hospital over time and the vertical axis is the mean \cd incidence in the network neighborhood of that hospital (the mean taken first over time and then over all network neighbors). We exclude hospitals with fewer than 100 patients from subsequent correlation analyses, leading to exclusion of 7.5\% (428) of all hospitals. The Pearson correlation coefficients are 0.47 and -0.01 for the low and high incidence regimes, respectively, which are separated by the vertical line.\label{Fig2}}
 \end{center}
\end{figure}

In our data, over the two-year period, there were a total of 313,214 \cd infections in the 5,677 hospitals included in the study (after all exclusion criteria were applied). We plot the mean \cd incidence for each hospital and the mean \cd incidence for its network neighbors in Fig.~\ref{Fig2}. We observe two distinct regimes, one for low \cd incidence and another for high \cd incidence. The incidence of the pathogen in a given hospital appears to be correlated with the incidence of the pathogen in its network neighbors as long as the incidence at the focal hospital is relatively low; this correlation appears to vanish for hospitals displaying higher \cd incidence. One possible explanation for this phenomenon is that, if there were only very few cases of \cd in the low incidence regime, the transfers of infected patients might go undetected, therefore inducing correlations among pathogen incidences across the network. Conversely, if pathogen incidence were high and local, such that hospital outbreaks are detected, patient transfers might be restructured to curb the further spread of the infection. We determine the boundary between the two regimes based on the strength of correlation in pathogen incidence and assign the value for the crossover between the two regimes (shown as the vertical line in Fig.~\ref{Fig2}). For \cd incidence below this threshold, the Pearson correlation coefficient $R \approx 0.47$ (95\% CI: 0.44, 0.49) whereas above the threshold  $R \approx -0.01$ (95\% CI: -0.08, 0.07), where the confidence intervals for the correlation coefficients where estimated using the Fisher $z$-transformation \cite{cite25}. This finding on the correlation of \cd incidence across hospitals that are neighbors in the transfer network supports the use of the transfer network as a substrate for the spread of nosocomial infections.

\subsection{Monitoring the system for hypothetical outbreaks}
We investigated the optimal selection and placement of network sensors for early detection of epidemics. We used three different strategies for selecting the sensor nodes based on their properties in the static network, choosing them based on their in-degree rank, out-degree rank, or choosing them at random. Nodes with a high in-degree are expected to be efficient at funneling in pathogens from their network environment, whereas nodes with a high out-degree are expected to rapidly funnel out their pathogens.

\begin{figure}
\begin{center}
 \includegraphics[width=8.6cm]{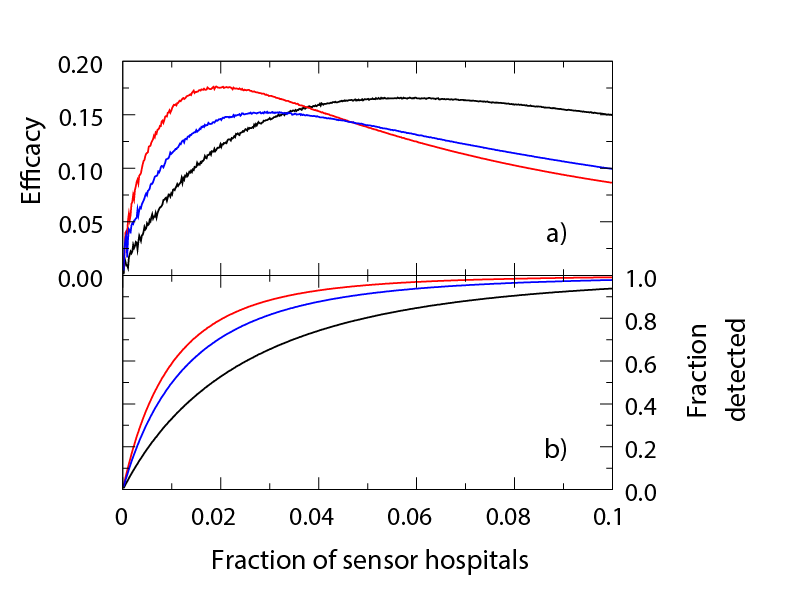}
 \caption{\textbf{Finding the optimal sensor set for the static implementation of the surveillance system.} Efficacy (a) and fraction of detected cases (b) on the static network as a function of the fraction of hospitals acting as sensors. The different curves represent different strategies for sensor selection: random selection (black), selection proportional to in-degree (red), and selection proportional to out-degree (blue). \label{Fig3}}
 \end{center}
\end{figure}

We implemented two versions of each strategy. In the \emph{static implementation}, the set of sensor hospitals was fixed in time, whereas in the \emph{dynamic implementation} different hospitals function as sensors at different times (see Methods). In Fig.~\ref{Fig3}, we show the results for the efficacy and the fraction of detected cases for the three strategies for the static implementation. The in-degree strategy achieves the highest efficacy with the lowest number of sensors and at most uses only 108 hospitals (1.9\% of all hospitals) as sensors. The out-degree strategy is second best and it uses 167 hospitals (2.9\%) as sensors. Both degree-based approaches outperform the random strategy that uses 332 hospitals (5.9\%). In terms of the fraction of detected cases, the three strategies perform similarly: 78\% for in-degree, 81\% for out-degree, and 84\% for the random strategy.

\begin{figure}
\begin{center}
 \includegraphics[width=8.6cm]{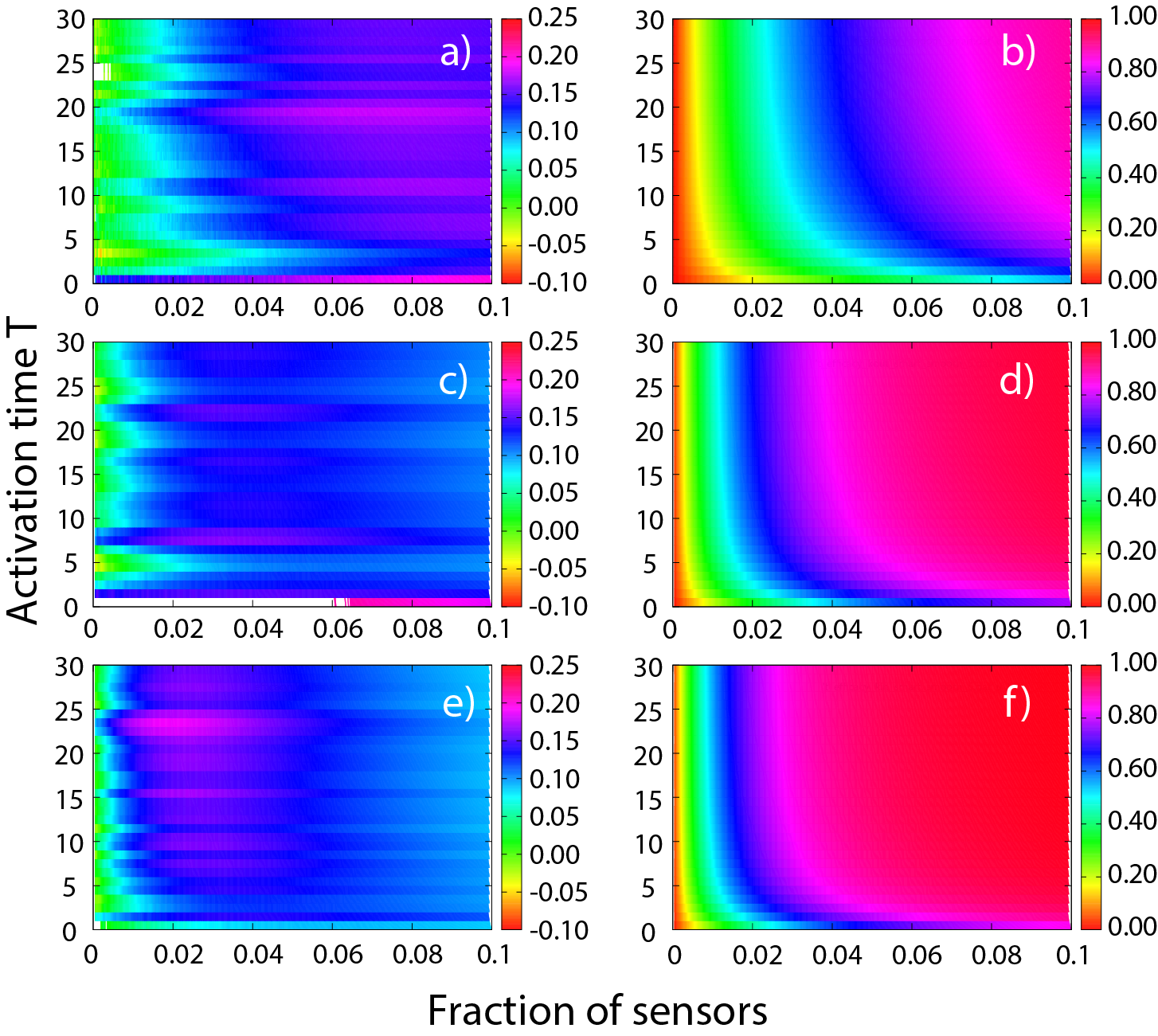}
 \caption{\textbf{Finding the optimal sensor set for the dynamic implementation of the surveillance system.} Heatmaps showing the efficacy (left column) and fraction of detected cases (right column) on the temporal transfer network. Results are shown as a function of the fraction of hospitals acting as sensors (horizontal axes) and the activity time that they implement (vertical axes). The rows of panels correspond to choosing the sensors randomly (top row), proportional to out-degree (middle row) and proportional to in-degree (bottom row).\label{Fig4}}
 \end{center}
\end{figure}

\begin{figure}
\begin{center}
 \includegraphics[width=8.6cm]{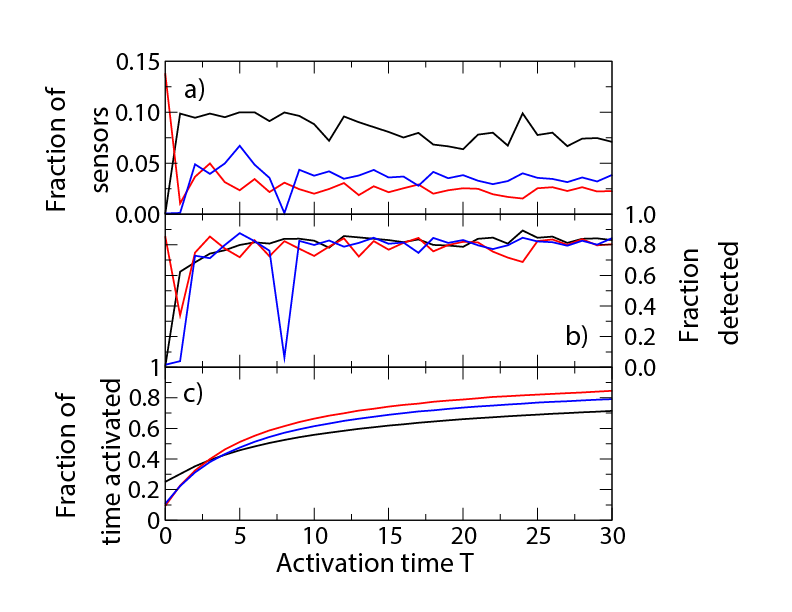}
 \caption{\textbf{Efficacy of temporal sensor sets.} \textbf{a)} Fraction of sensors for the most efficient sensor set from the temporal network for sensors chosen at random (black), proportional to in-degree (red), and proportional to out-degree (blue). We have smoothened the efficacy curves by averaging the results using a window of 5 sensors. \textbf{b)} Fraction of detected cases for the most efficient sensor set. \textbf{c)} Average fraction of time that a sensor stays in the active state (same color code as on the left).\label{Fig5}}
 \end{center}
\end{figure}

In Fig.~\ref{Fig4}, we show the efficacy and fraction of detected cases for the three strategies for the dynamic implementation as a function of the number of sensors and the \emph{activation time} $T$, the length of the time period that the hospital will be incorporated in the sensor set upon admitting a \cd patient. Except for very low activation times of the order of a few days, the measures of efficacy and fraction of detected cases are almost unaffected by this parameter. As can be seen in Fig.~\ref{Fig5}, the optimal sensor set of a strategy stabilizes after $T=5$ days. These results corroborate the finding that choosing sensors based on in-degree is the best overall strategy, followed by out-degree, and then the random strategy. All of the strategies result in similar sizes for the most efficient sensor sets as in the static case. In terms of the fraction of detected cases, all three strategies perform similarly, each covering about 80\% of the cases. We find that the average time a sensor spends in the active state increases as a function of the activation time $T$. Therefore, an optimal approach is to choose the smallest activation time $T$ that does not deteriorate performance of the sensor system in terms of the fraction of detected cases. For an activation time $T=5$, the average fraction of time sensors spend in the active state is $0.51$ for in-degree based selection, $0.47$ for out-degree based selection, and $0.46$ for the random strategy.

\begin{figure}
\begin{center}
 \includegraphics[width=8.6cm]{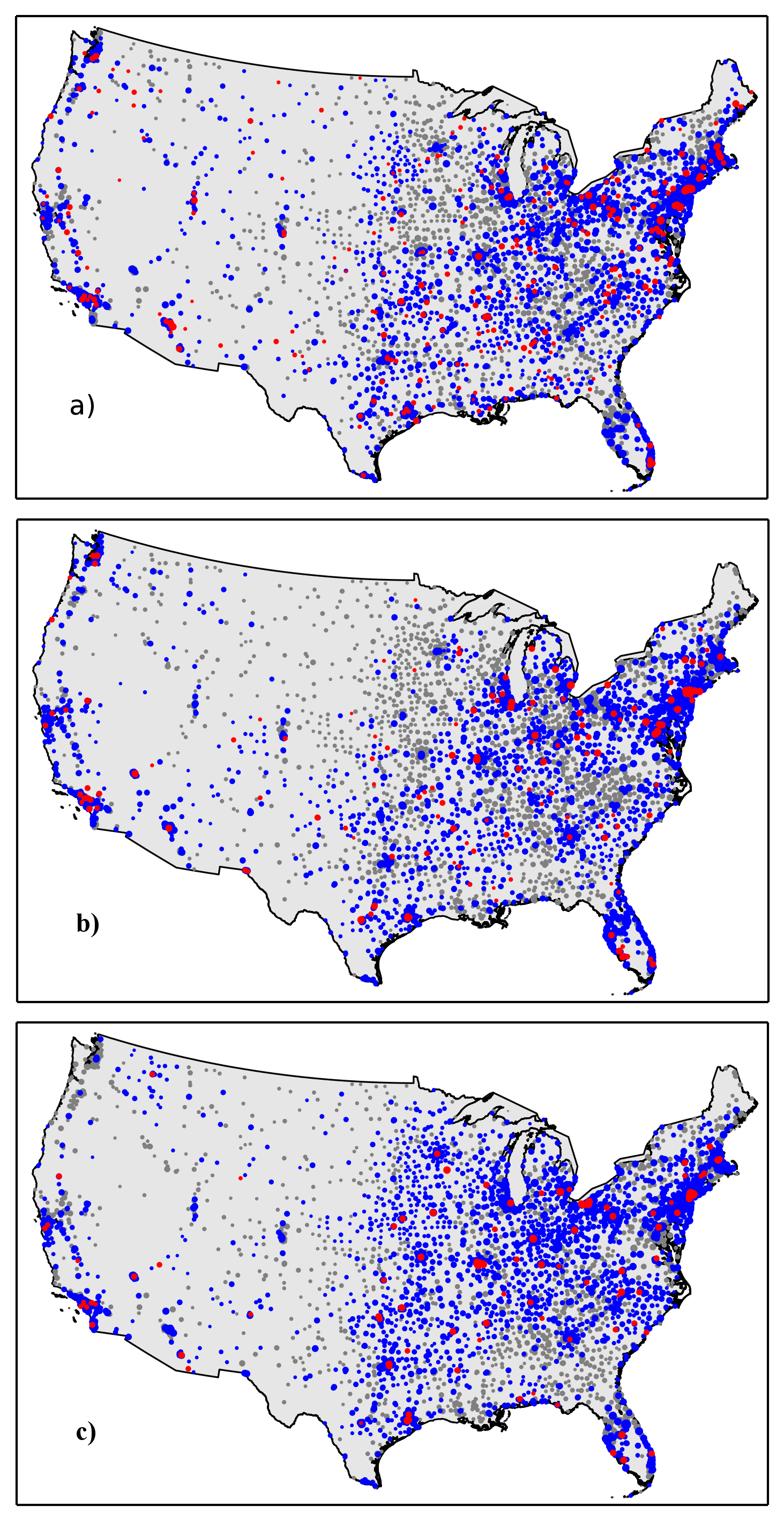}
 \caption{\textbf{Spatial locations of optimal sensor sets} in the static transfer network based on in-degree (a), out-degree (b), random (c). In red are the sensor hospitals, in blue their first neighbors and in grey those uncovered by the sensor set.\label{Fig6}}
 \end{center}
\end{figure}

In Fig.~\ref{Fig6} an instance for the optimal sensor set derived from each strategy in the static implementation is plotted in a map. Sensor hospitals are plotted in red, while their first neighbors in blue and the rest in grey. Their size encodes the number of \cd cases they host in the full study period. We visually see that the number of blue and red hospitals are more or less similar for all strategies, while the number of sensor hospitals (in red) decreases from the random (Fig.~\ref{Fig6}a), to the out-degree (Fig.~\ref{Fig6}b), to the in-degree strategy (Fig.~\ref{Fig6}c).

\section{CONCLUSIONS}

We studied a network defined by the transfer of 12.5M Medicare patients across 5,667 US hospitals over a 2-year period. We found the network to be strongly geographically embedded, with 90\% of all transfers spanning a distance less than 200km. We found that the transfer network could plausibly be used as a substrate for the spread of pathogens: we observed a positive correlation for \cd incidence between hospitals and their network neighbors, identifying two qualitatively distinct regimes corresponding to low and high \cd incidence. Finally, we showed that selecting hospitals as sensors based on their in-degree in the static network was able to detect a large fraction of infections. Furthermore, an activation time of just 5 to 7 days using the dynamic sensor implementation is sufficient to achieve this surveillance with just 2\% of the hospitals acting as sensors. These results support our conceptual model that the structure of the nationwide hospital patient transfer network is important for the spread of health-care associated infections, likely well beyond the illustrative case of C. diff considered here. In particular, our work highlights the need to monitor the network of transfersâ not just individual hospitalsâ in order to monitor infectious outbreaks. 

It is possible that other sorts of pathogens might need a different number of sensor hospitals, a different set of sensor hospitals, or different surveillance windows. Nevertheless, it is clear that the health of the entire hospital system, from the perspective of nosocomial infections or other outbreaks, could be monitored by leveraging the network structure of patient transfers.  

Our study has several limitations. First, the data we used to map the hospital networks are from 2006 and 2007. However, given that hospital transfer patterns are strongly embedded in the geography of the country, as we also demonstrated here, we do not expect the age of the data to affect our results substantially. Second, we cannot assess the extent to which unobserved policies or commercial constraints might have affected the flow of patients from one hospital to another; however, these policies merely affected patient transfers, which are, in any case, observable in the current and similar future data. Third, our analyses and models assume that patient transfers are the only mechanism responsible for the spread of infections. There are, of course, other vectors or means that might result in hospitals being infected, such as the movement of physicians, nurses, and other health care staff between hospitals. Finally, in this analysis, we did not make use of the fine-scale temporal information available in transfer data; future work could evaluate how bursts of infected patients, perhaps on particular days of the week, might contribute to an epidemic.

Understanding the structure and dynamics of the hospital transfer network for the spread of real infections has a number of important implications. Empirical data could be used, either periodically or perhaps even in real timeâ to map networks of patient movement in the US health care system, and this network could then be used monitor the spread of nosocomial and other infections in the network.  In our estimation, such a system could detect 80\% of \cd cases using just 2\% of hospitals as network sensors. Our methods suggest practicable strategies for identifying which hospitals should serve a surveillance function for the whole system and, in the dynamic implementation, how long the sensors should retain a higher level of alertness after each index case. These tools would be useful not only for public health interventions in the case of natural epidemics, but also in the case of deliberate ones, such as those due to a possible bioterror attack. In conclusion, the actual structure and flow pattern of patients across US hospitals confers certain specific vulnerabilities and defenses, regardless of the biology of the pathogen per se, placing theoretical bounds on any effective containment strategy directed at a contagious pathogen.

\section{MATERIALS AND METHODS}

\subsection{Study data}
We study hospital-to-hospital transfer patterns of the entire population of US Medicare patients over a two-year period. Medicare provides almost universal coverage to all Americans aged 65 and older, about 15\% of the US population \cite{cite17}; and about 37\% of all hospital admissions in 2003 were for Medicare patients \cite{cite18}. We used a 100\% sample of the Medicare Provider Analysis and Review (MedPAR) files for calendar years 2006 and 2007. The MedPAR files contain diagnosis, procedure, and billing information on all inpatient and skilled nursing facility (SNF) stays. Our study cohort consisted of Medicare patients aged 65 or older with a hospital stay at an acute medical or surgical hospital with an active record in the American Hospital Association (AHA) 2005 database \cite{cite19}. Before applying these exclusion criteria, we identified 26.4 million stays of 12.5 million patients in 6,278 different hospitals. After the exclusions, our final cohort consisted of 21.0 million inpatient stays of 10.4 million patients in 5,667 different hospitals. 

\subsection{Hospital-to-hospital transfers}
According to our definition, a hospital-to-hospital transfer occurs whenever a patient is discharged from one hospital and admitted to another hospital on the same calendar day. Note that a minority of transfers as defined here may not correspond to actual formal transfers of patients. For example, a patient could be discharged from hospital A and then be re-admitted to hospital B on the same day for a reason that is unrelated to her stay at hospital A. From an epidemiological point of view, however, these are essentially equivalent to formal patient transfers. (Our results change little if we relax the definition of hospital transfers to allow the re-admission to take place the day following the day of discharge. See the appendices.) Using this definition of transfer, we identified 936,101 transfer events taking place between 76,003 pairs of hospitals. 

\subsection{Constructing the transfer network}
We consider a network representation of the patient transfers across hospitals. In this framework, hospitals are represented as nodes and a transfer of a total of $x$ patients on day $d$ from hospital $i$ to hospital $j$ is represented as a directed edge from node $i$ to node $j$ with weight $x$ on day $d$. The longitudinal sequence of patient transfers forms a directed, weighted, temporal network. We consider a static representation of the network that retains no temporal information of patient transfers by aggregating the data for the two-year period, where the weight of the edge from node $i$ to node $j$ is the mean daily number of patient transfers through that edge, i.e., the total number of transfers from hospital $i$ to hospital $j$ during the study period divided by the number of days in the period (730). 

\subsection{C.~Diff.~incidence on the transfer network}
The transfer of infected patients from one hospital to another can result in pathogen transmission between them. Given that the MedPAR files contain diagnosis codes for each patient, we investigated the incidence of \emph{Clostridium difficile} (\emph{C. diff.}) infections and its correlation with properties of the transfer network. \cd is an anaerobic, gram-positive, spore-forming bacteria that occurs frequently in health care settings. It is found in over 20\% of patients who have been hospitalized for more than one week. The disease is spread by ingestion of \cd spores, which are very hardy and can persist on environmental surfaces for months without proper hygiene \cite{cite20}. \cd associated infections kill an estimated 14,000 people a year in the US as a result of institutional infections \cite{cite21}. We ascertained incident cases of \cd infection by identifying any hospital admissions with ICD-9 diagnostic code 008.45. The sensitivity and specificity of using ICD-9 codes to identify \cd infections have been reported by multiple groups to be adequate for identifying overall \cd burden for epidemiological purposes \cite{cite22,cite23,cite24}. Given the relative \cd incidence at each hospital, defined as the fraction of patients with that particular diagnosis over the study period,  we plot the average relative \cd incidence in the neighborhood of each hospital against its own \cd incidence in Figure ~\ref{Fig2}. We quantify the correlation using the Pearson linear correlation coefficient.
 
\subsection{Sensor placement on the hospital network}
It might be possible to make use of the properties of the hospital-hospital transfer network to set up a real-time surveillance system for infections, such as a new strain of antibiotic-resistant \cd For this application, it is unlikely that exhaustive data would be available for all hospitals all the time, and this limitation calls for a parsimonious approach where only a subset of hospitals needs to be monitored at any given time. We call these monitored hospitals ``network sensors'' in the sense that they could be used to sense incipient epidemics. We consider three different prescriptions for sensor placement: (1) choose sensor hospitals in proportion to their in-degree rank in the static network; (2) choose sensor hospitals in proportion to their out-degree rank in the static network; and (3) choose sensor hospitals uniformly at random from the set of all hospitals. In our simulations, we assume that a monitored hospital is able to detect every infected patient who is present either in the hospital itself or in any of its network neighbors to which it is connected via patient transfers. While this assumption is made primarily for methodological reasons and may not hold in practice, the relative performance (the ordering) of the three prescriptions for selecting sensors remains unaffected if the assumption were relaxed. To learn about the potential of the hospital sensor framework to detect epidemics, we investigate its best-case performance by determining the optimal sensor set for the observed data (see appendices). We expect that its performance would be somewhat reduced for an independent test data set (data not used as part of the training of the method).

\subsection{Determining the optimal sensor set}
We define the relative efficacy of the sensor $E_N$ set as
\begin{equation}
 E_N=\frac{D_N}{ND_1}-\frac{M-D_N}{M}
\end{equation}
where $N$ is the number of sensors in the sensor set, $D_N$ the number of infected patients detected by a sensor set of $N$ sensors, and $M$ is the total number of \cd cases in the network. While adding sensors to the system always improves its overall performance, any sensor set exhibits diminishing marginal returns in the sense that the per-sensor increment in performance declines with each added sensor. The first term in the definition corresponds to the number of detected cases normalized by the number of cases that would be detected if all sensors were as efficacious as the first sensor in the sensor set. The second term is a penalty term that corresponds to the fraction of undetected cases. High relative efficacy is therefore a combination of selecting a set of sensors that are as close as possible to the efficaciousness of the first sensor in the set and having these sensors miss as small a proportion of cases as possible. Note that the two terms in the definition of the relative efficacy could be assigned different weights; however, here, we opted for the simplest approach and only ensured that the two contributions are measured on the same scale.

\subsection{Static and dynamic implementation of network sensors}
We implement the sensor framework in two different ways. In the static implementation, the sensors are always active, whereas, in the dynamic implementation, the sensors are either passive or active. When a sensor is passive, it can only detect infections in the hospital itself. Whenever an infection is detected, the sensor either transitions from the passive state to the active state for a period of $T$ days or, if already in the active state, remains in that state for another $T$ days. In addition to the efficacy of the sensor sets, for both implementations, we keep track of the fraction of \cd cases that are detected in order to assess the performance of the sensor system.

\noindent \textbf{Static implementation} Since we know the number of \cd cases in each hospital at any given time, we simply count the number of cases in the sensor hospitals and their network neighbors. We average the results by generating 10,000 independent realizations of sensor sets for each of the three different prescriptions of choosing sensors (in-degree, out-degree, random). The optimal sensor set for each strategy is the one with maximum efficacy.

\noindent \textbf{Dynamic implementation} We monitor the admission times of \cd patients at each hospital, and whenever such a patient is admitted, we incorporate the hospital in the sensor set for $T$ days following the admission, a time period we call the activation time. Once added to the sensor set, the hospital can detect the \cd cases present in the hospital itself and its network neighbors for a total of $T$ days. The efficacy of the sensor system therefore depends on the value of $T$, and we compute the efficacy of the sensors for $T$ from 0 to 100 days (shown from 0 to 30 days in Fig.~\ref{Fig5}). For each combination of parameter values, the number of sensors and the activation time, and for each strategy of prescribing sensors, we perform 1,000 independent realizations of the sensor selection process. We also track the average time each sensor stays in the active state. An optimal sensor set is one that has maximal efficacy for activation time $T$, minimizes the average time the sensors stay active, and maximizes the fraction of detected cases.

\begin{acknowledgments}
We thank Laurie Meneades for the expert assistance required to build the dataset. JFG and JPO are joint first authors of this article. 
\end{acknowledgments}

\renewcommand{\thefigure}{A\arabic{figure}}
\setcounter{figure}{0}
\newpage

\section{APPENDICES}

\subsection{Transfer network}

\begin{figure}[H]
\begin{center}
 \includegraphics[width=8.6cm]{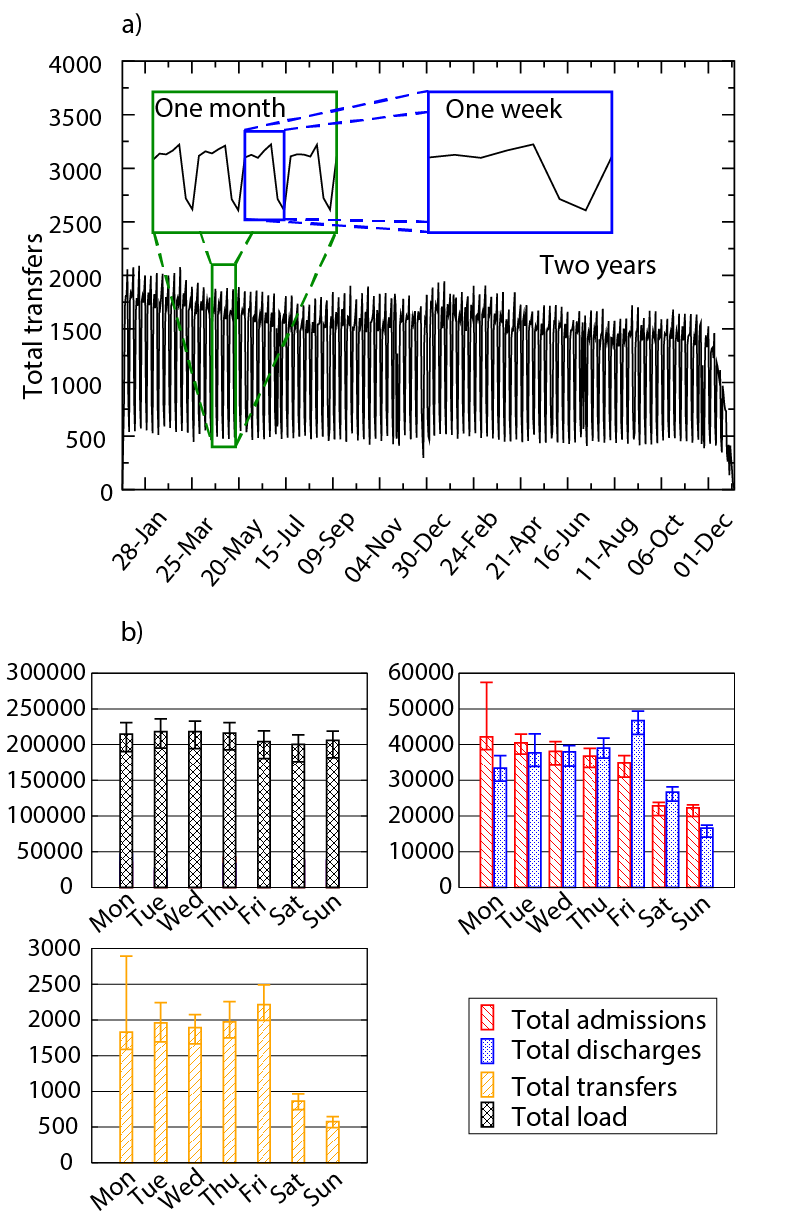}
 \caption{\textbf{ Hospital Transfer Network HTN.} \textbf{a)} Total number of transfers in the system  as a function of time for the two years of data. We can see seasonal and weekly oscillations. \textbf{b)} Median, 5- and 95-percentile for several quantities of interest for the different days of the week.\label{FigS1}}
 \end{center}
\end{figure}

We characterize the temporal nature of hospital usage by showing the time series of the number of transfers in Figure \ref{FigS1} (a). A clear seasonal oscillation is visible, and at a finer temporal scale, a weekly periodic cycle is also observable, where Saturdays and Sundays are the least active days of the week and Mondays the most active and also the most variable. In Figure \ref{FigS1} (b) there are periodic oscillations in many of the quantities of interest, such as the number of patients staying overnight at hospitals, number of admissions, discharges, and transfers.

\begin{figure}[H]
\begin{center}
 \includegraphics[width=8.6cm]{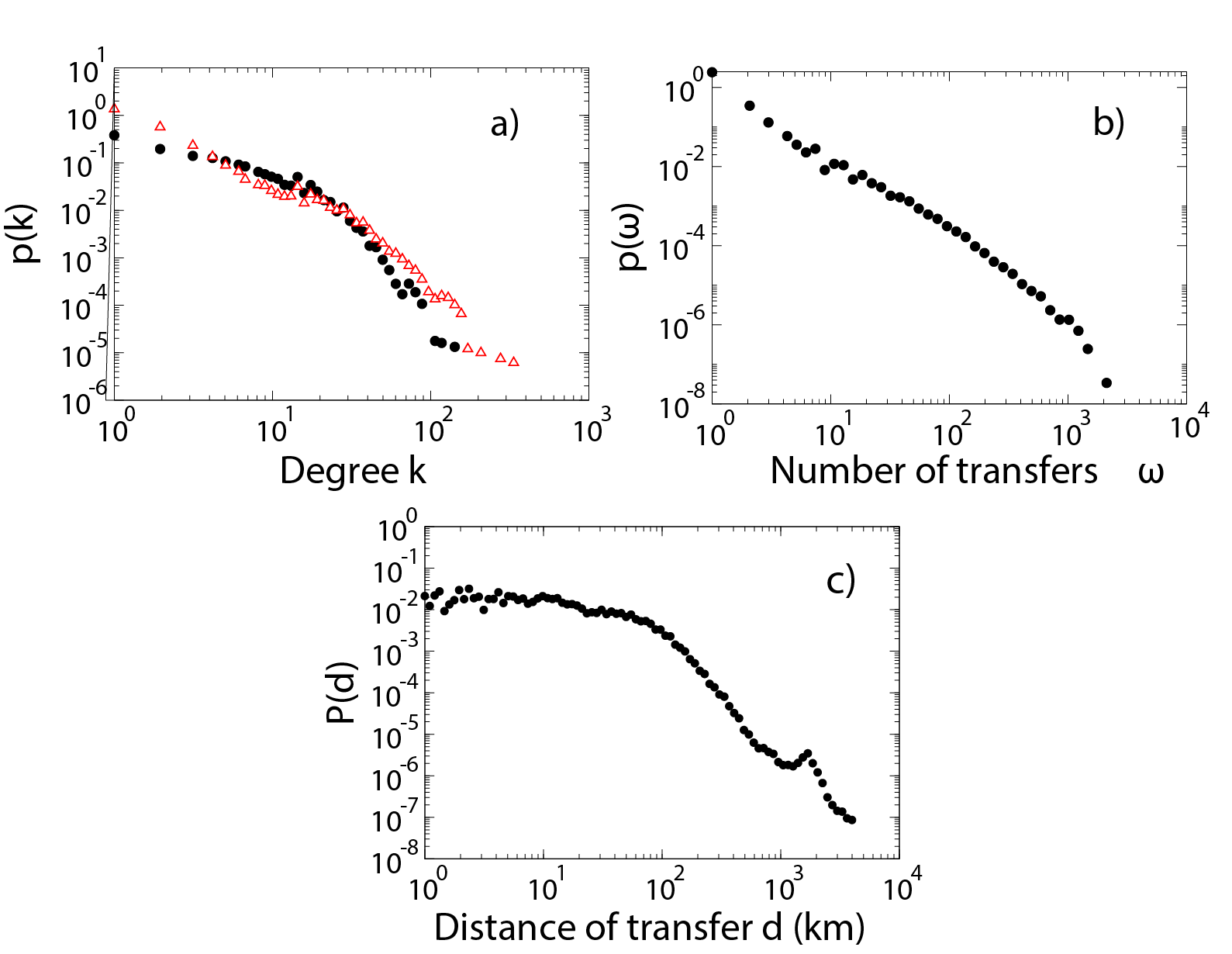}
 \caption{\textbf{ Topological and geographical characteristics of the transfer network.} \textbf{a)} Distributions for in- (red open triangles) and out-degree (black solid circles). \textbf{b)} Distribution for the number of transfers per connection $\omega$. \textbf{c)} Transfer distance distribution. \label{FigS2}}
 \end{center}
\end{figure}

We then examine the structural connectivity and geographic characteristics of the static transfer network (see Fig. \ref{FigS2}). In terms of network topology, the in-degree distribution has a broader tail than the out-degree distribution. The network has an average (local) clustering coefficient of 0.51. This coefficient measures the probability that any two hospitals connected to an index hospital are in turn connected to each other, forming a closed triad (a cycle of three nodes and three edges). A random graph with the same number of nodes and edges yields an average local clustering coefficient of 0.0057$\pm$ 0.0001 (SE), which is substantially lower than the observed value, a finding that likely reflects the network’s geographic embeddedness. The average shortest path length of the network is 4.69. To put this number in perspective, we performed network randomizations using a slight variant of the directed configuration model that preserves both in-degree and out-degree distributions \cite{citeS1}. This approach gave rise to an average shortest path length of 3.6 $\pm$ 0.4 (SE). The observed network is therefore a somewhat “larger world” than what would be expected by chance, but this is almost certainly driven by the underlying geography and the objective of keeping transfers as short as possible. In fact, about 90\% of the transfers are to hospitals less than 200km away.

Degree assortativity is the concept that nodes with many connections tend to be connected to other nodes with many connections \cite{citeS2,citeS3}. When the static network is taken as undirected, we can use the assortativity coefficient to measure the extent to which the degrees of hospitals in each pair of connected hospitals are similar. We obtain a slightly negative value of -0.06, but similar values of -0.005 $\pm$ 0.001 (SE) also arise from randomizations of the network using the algorithm discussed above. Consequently, there is no statistically significant assortativity in the network over and above what would be expected by chance given the network’s degree distributions.

\subsection{Robustness of the transfer extraction}

\begin{figure}[H]
\begin{center}
 \includegraphics[width=8.6cm]{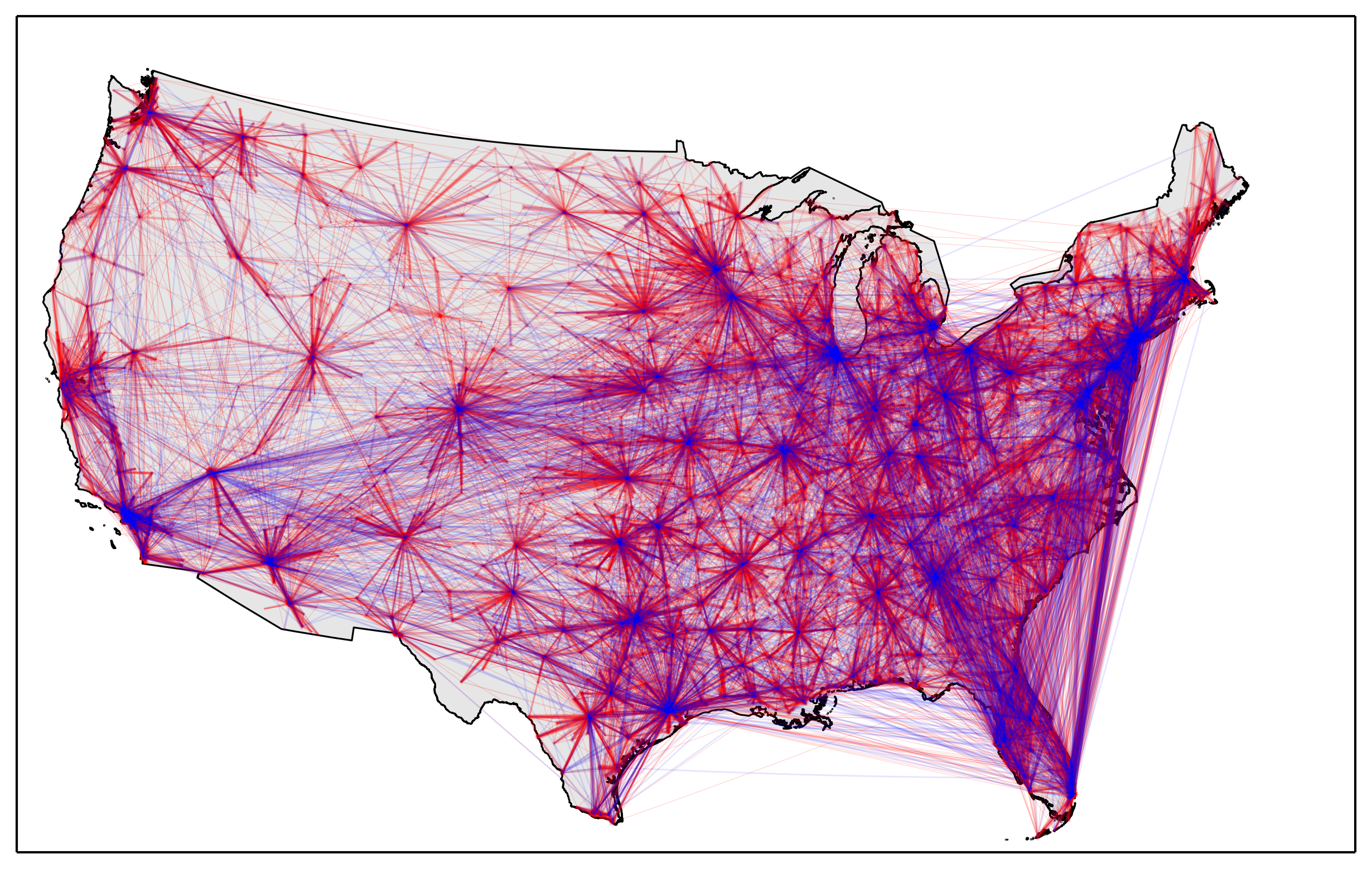}
 \caption{\textbf{ Comparison of the transfer network based on the 1-day and 2-day rules.} The network is constructed by aggretating transfer data over the full two-year period. Red edges correspond to the connections induced by the 1-day rule and the blue edges correspond to the additional edges that appear when considering the 2-day rule.\label{FigS3}}
 \end{center}
\end{figure}

\begin{figure}[H]
\begin{center}
 \includegraphics[width=8.6cm]{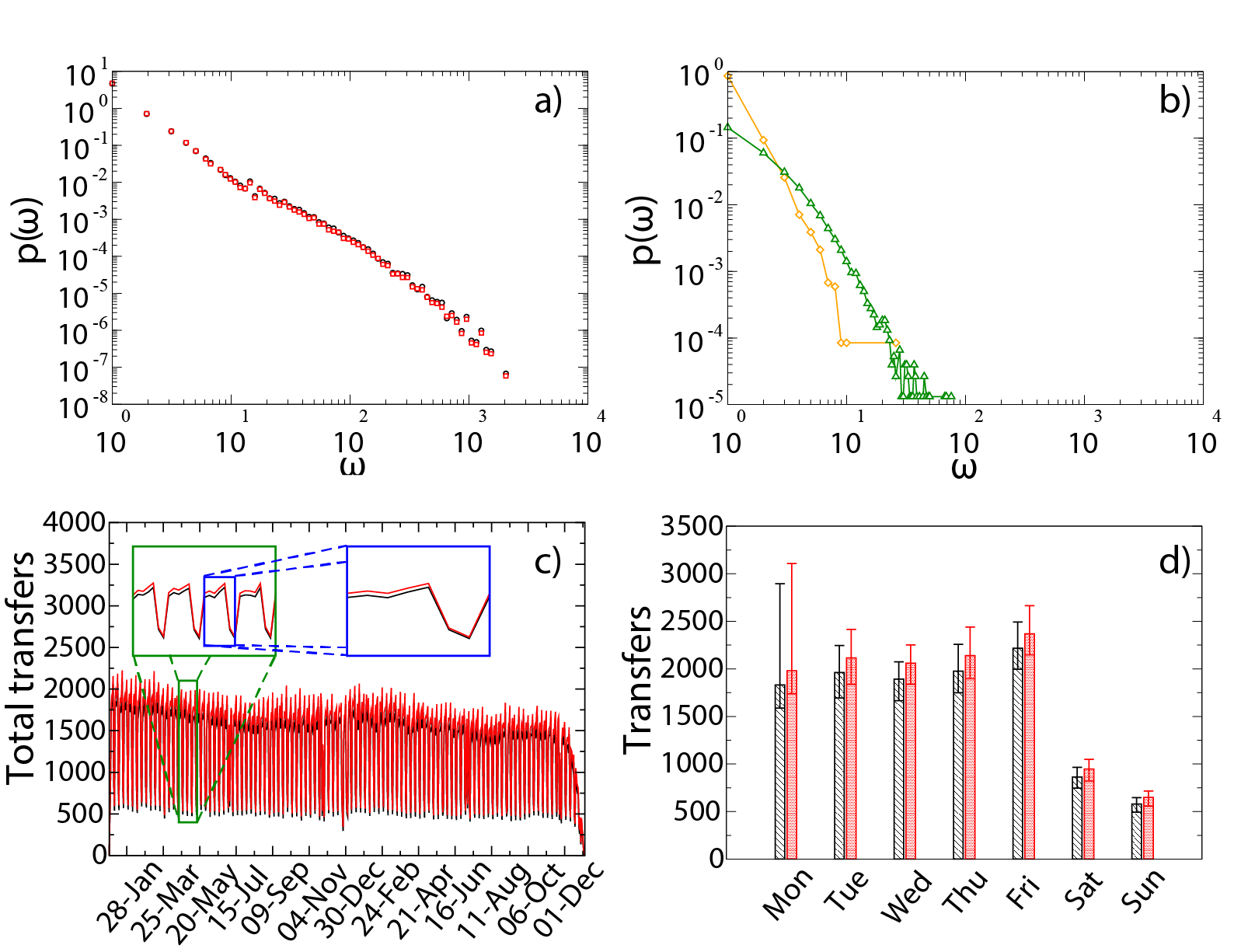}
 \caption{\textbf{ Comparison of transfer window of one and two days.} \textbf{a)} Distributions of the number of transfers per connection ω in black for one day transfers (1-day rule) and in red for one or two day transfers (2-day rule).  \textbf{b)} Distribution of the number of transfers per connection for the edges that appear when using the 2-day rule. Two-day transfers (orange diamonds) and of the difference in the number of transfers for the connections that are shared by the two rules (green triangles). \textbf{c)} Temporal evolution of the total number of transfers for one day and two day transfers. The insets show a four-week and a one-week window, showing the periodicities in the data. \textbf{d)} Median, 5- and 95 percentiles for the transfers aggregated by day of the week. Again a comparison of one day and two day transfers demonstrates that they are qualitatively very similar.\label{FigS4}}
 \end{center}
\end{figure}

Since the patient transfers are not explicit in the data but instead need to be inferred from the data, we investigated the robustness of some of the results to our definition of what constitutes a hospital transfer. Instead of requiring readmission on the day of discharge, we relaxed this definition by allowing the readmission to take place also on the day after discharge. A visual examination of Fig. \ref{FigS3} shows that the edges induced by the same-day rule (red edges) and the additional edges that result using the relaxed rule (blue edges). This relaxation leads to 67472 additional transfers (7.2\% increase). There are 11827 new edges that appear on the transfer network (15.6\% increase), with an average transfer load of 1.2 with a standard deviation of 0.7. For the connections that appear under both rules, the difference in transfer loads averages to 0.7 transfers with a standard deviation of 1.9. The distribution of edge weights for both cases are shown in the upper left panel of Fig. \ref{FigS4}, and the two distributions appear visually very similar to one another. The weight distribution of the additional edges, as well as the distribution of weight differences for the common edges in both cases can be seen in the upper right panel of Fig. \ref{FigS4}. The range of this distribution is much more constrained than that of the actual weight distributions. The number of transfers increases, but the patterns remain essentially the same both temporally and topologically. For the temporal patterns, see the lower panels of Fig. \ref{FigS4}. Note also that both measures of transfers are strictly speaking wrong, as the first one based on the one-day rule is really a lower bound on the number of transfers and the second one (based on the relaxed rule) is an upper bound. Given the similarity of these findings across the two rules, in the following we work with the lower bound (same day discharge and readmission).

\subsection{Optimal sensor set}

We determine the best sensor set we could have possibly chosen given the observed data. In order to do this, we use greedy algorithms \cite{citeS4} as checking all possible combinations of hospitals to use as sensors grows exponentially in the number of hospitals and is therefore not feasible for any but the smallest hospital transfer networks. For a fast algorithm that is not guaranteed to give the optimal answer (as is true with any heuristic algorithm), we choose the sensors sequentially. We first compute the number of cases each hospital would detect and we choose the one that will detect the highest number of cases. We then re-compute how many new cases would be covered by each subsequent hospital if added to the existing sensor set. This continues until we find the sensor set that covers all cases. As mentioned above, this procedure does not guarantee that we will choose the optimal sensor set given a number of sensors N, but it is however very efficient and yields an effective sensor set not far from the optimal one. In order to check that our solution is sufficiently close to the actual best solution, we used simulated annealing \cite{citeS5}. The simulated annealing procedure is suitable for optimization problems of large scale, especially ones where a desired global extremum is hidden among many, poorer, local extrema. There is an objective function to be minimized, in our case the coverage of cases to be maximized, but the space over which that function is defined is not simply the N-dimensional space of N continuously variable parameters. Rather, it is a discrete, but very large, configuration space with the number of elements factorially large, so that they cannot be explored exhaustively. This result is in agreement with the result of the fast sequential algorithm.

\begin{figure}[H]
\begin{center}
 \includegraphics[width=8.6cm]{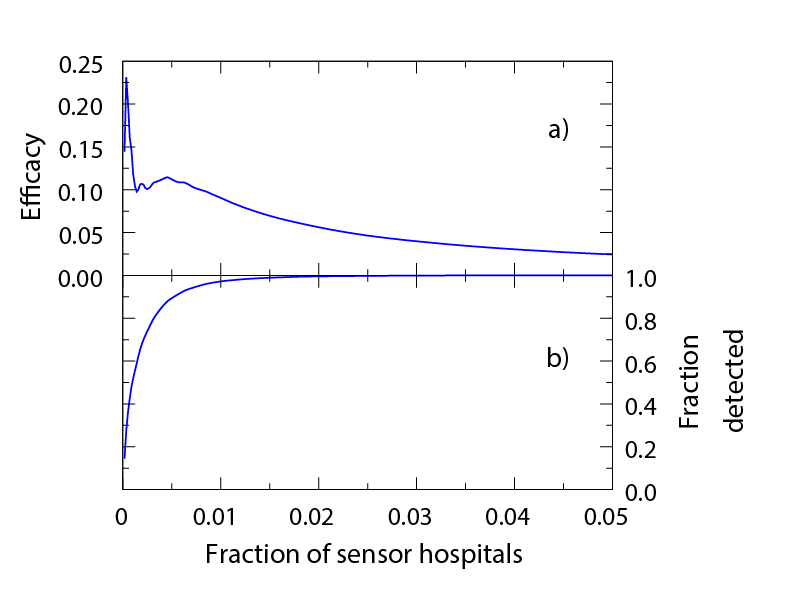}
 \caption{\textbf{ Finding the optimal number of sensors for the best sensor selection (static network).} \textbf{a)} shows the efficacy and \textbf{b)} the fraction of detected cases, both as a function of the fraction of hospitals used as sensors. There is a peak for a very low fraction of sensors, but this point however corresponds to no more than 30\% of detected cases. The second peak located at around 0.005 (using 0.5\% of hospitals as sensors) is able to detect over 80\% of the cases. \label{FigS5}}
 \end{center}
\end{figure}

In Fig. \ref{FigS5} we show the results of finding the sensor set that maximizes the number of detected cases in the training dataset for the static network case. This method is data-based and tries to maximize the number of detected cases without the use of any strategy of choosing sensors  other than the optimization procedure. In this case we find that for a very small number of 26 (0.46\%) sensors, we can detect 88\% of the cases. This very high performance is however likely a consequence of over-fitting the model to the observed (training) data. Using this set of hospitals as sensors for a new dataset on patient transfers would likely result in lower (and more variable) performance of the sensor system.

\begin{figure}[H]
\begin{center}
 \includegraphics[width=8.6cm]{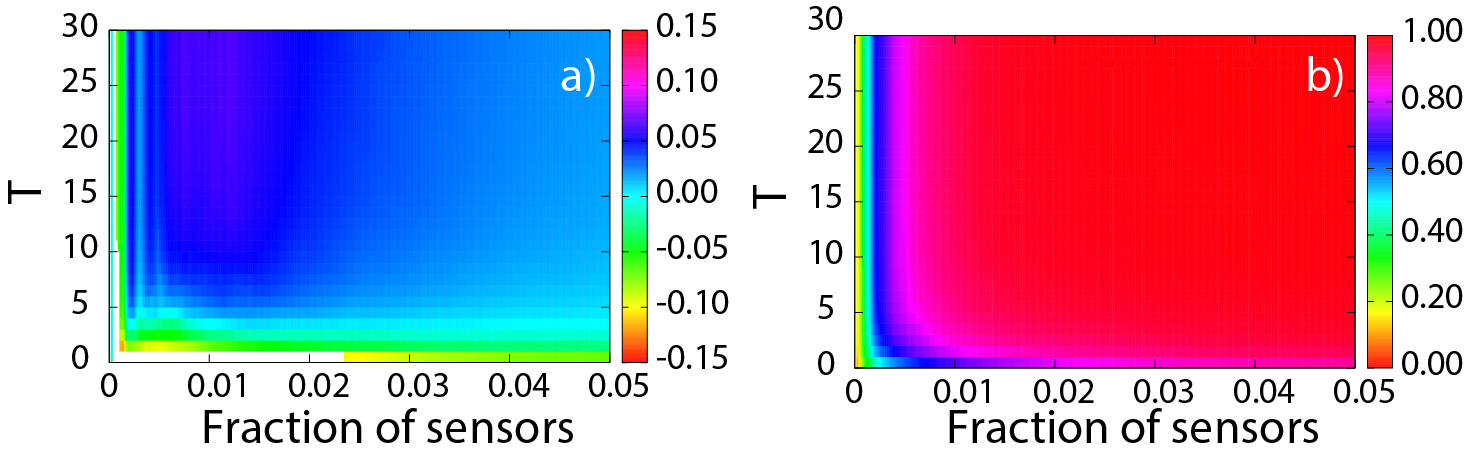}
 \caption{\textbf{ Finding the optimal number of sensors for the best sensor selection (dynamic network).} Efficacy (a) and the fraction of detected cases (b) as a function of the fraction of sensors and the activation time $T$. \label{FigS6}}
 \end{center}
\end{figure}

In Fig. \ref{FigS6} we can see the results of performing the same analysis for the dynamic implementation. Now the hospitals that are sensors act only as sensors for a period T days after admitting a patient with a C.diff infection. The greedy method for choosing sensors works as in the static case, but now taking into account the temporal restrictions for the cases that the sensor system is able to detect. The result is similar to the results of the other methods when moving from the static to the dynamic case. The results are different for a small value of the activation time, below one week, but remain basically unchanged as the activation time is raised.

\begin{figure}[H]
\begin{center}
 \includegraphics[width=8.6cm]{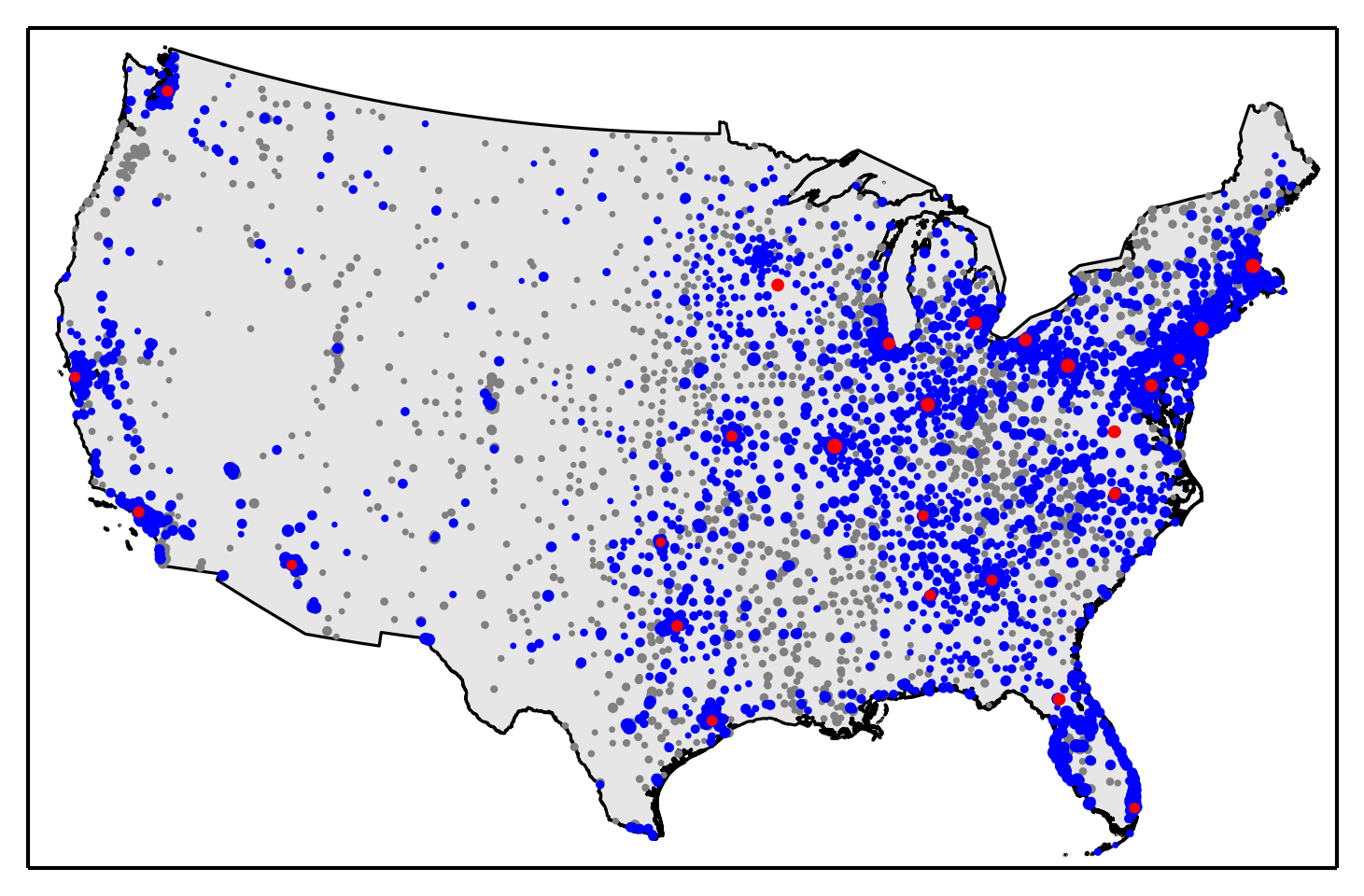}
 \caption{\textbf{ Spatial positioning of the optimal sensor set.} Red dots represent the sensor hospitals and blue dots are (nearest) neighbors of sensor hospitals. The size of each dot represents the mean C. diff incidence taken over the 2-year period at the hospital.\label{FigS7}}
 \end{center}
\end{figure}

Finally in Fig. \ref{FigS7} we can see the sensor set that is the result of the optimization for the aggregated case.

\subsection{Robustness of sensor set performance}

Thte performance of statistical methods is generally quantified using some error metric, and most fitting procedures attempt to minimize this error in the process of finding suitable values for model parameters. It is often possible to reduce this training error by increasing model complexity, but generally the goal of modeling is to have the model perform well on a test data set, ideally an independent data set, that the model was not trained on. Good performance on a test data set, quantified by a low test error, generally leads to better overall model performance and avoids the problem of over fitting, which refers to the model adapting to the test data “too well” at the expense of poor generalizability to different realizations of data from the same data generating mechanism.

In analogy with this approach to statistical learning, we performed a series of analyses to investigate the performance of sensor sets derived from one set of data and tested on another. The objective of the analysis is twofold. First, it will enable us to ascertain the validity of our methods when applied to training data, i.e., data not used to select the set of sensors. Two, given that there are likely temporal correlations in the data, it enables us to study the performance of sensor sets on data that are temporally far removed from the training data.

Here we divided our data to disjoint (non-overlapping) windows of width L, where we used values of 1 month, 2 months, 4 months, 6 months, and a year for L. For any given window, we take the first window to be our training data and use all subsequent windows as different realizations of test data. We used the training data for generating the sensor sets (based on in-degree, out-degree, and the greedy algorithm; we exclude considerations of the random stragy here because there is no real distinction between testing and training) and evaluated the relative efficacy and the percentage of cases detected separately for each test data window.

\begin{figure}
\begin{center}
 \includegraphics[width=8.6cm]{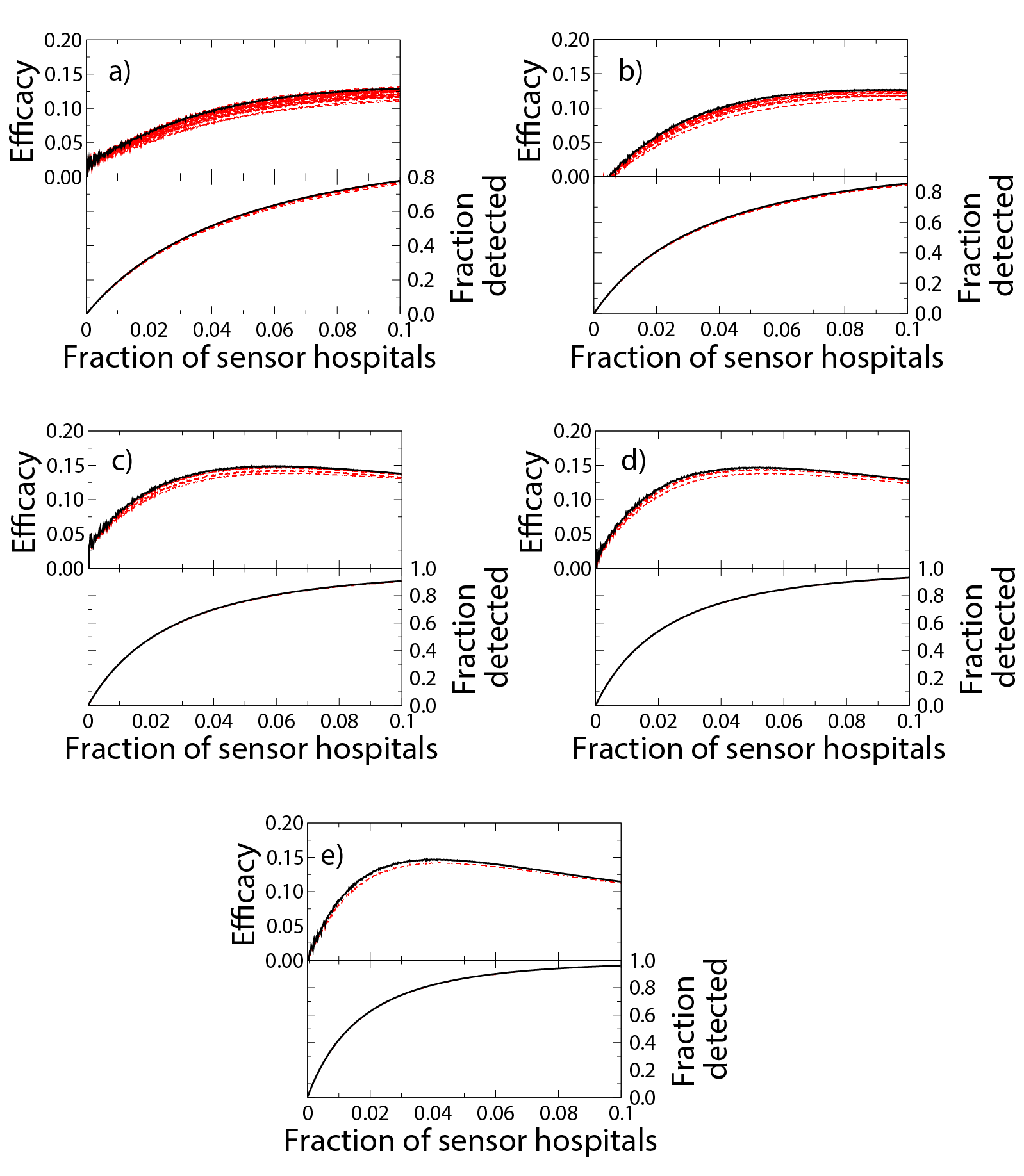}
 \caption{\textbf{ Out-degree strategy: training vs. test data.} Due to temporal correlations in the data, the sensor sets derived from the first slice of data perform comparably to their performance on the training set when applied to the remaining slices of data as test data. In all the plots, the results for the training set are shown as black solid lines while the red dashed lines refer to the sensor set applied to the test data sets. From left to right and top to bottom, the different plots refer to window widths of 1 (a), 2 (b), 4 (c), 6 (d), and 12 (e) months.\label{FigS8}}
 \end{center}
\end{figure}

\begin{figure}
\begin{center}
 \includegraphics[width=8.6cm]{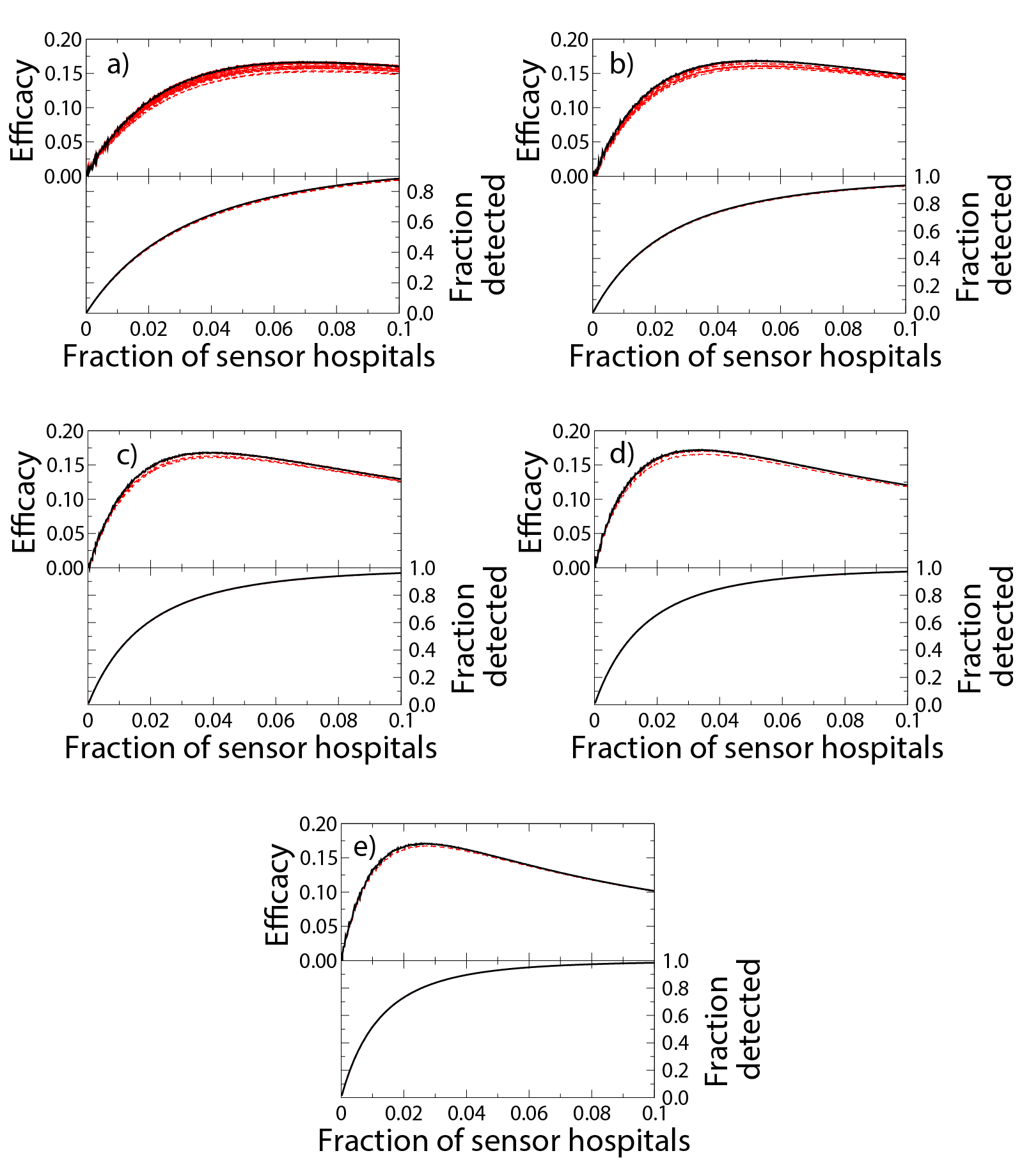}
 \caption{\textbf{ In-degree strategy: information vs. validation sets.} The panels are arranged as above. Due to the temporal correlation of the data the sensor sets derived from the first slice of data perform comparably to their performance on the training set when applied to the remaining slices of data as test data.\label{FigS9}}
 \end{center}
\end{figure}

\begin{figure}
\begin{center}
 \includegraphics[width=8.6cm]{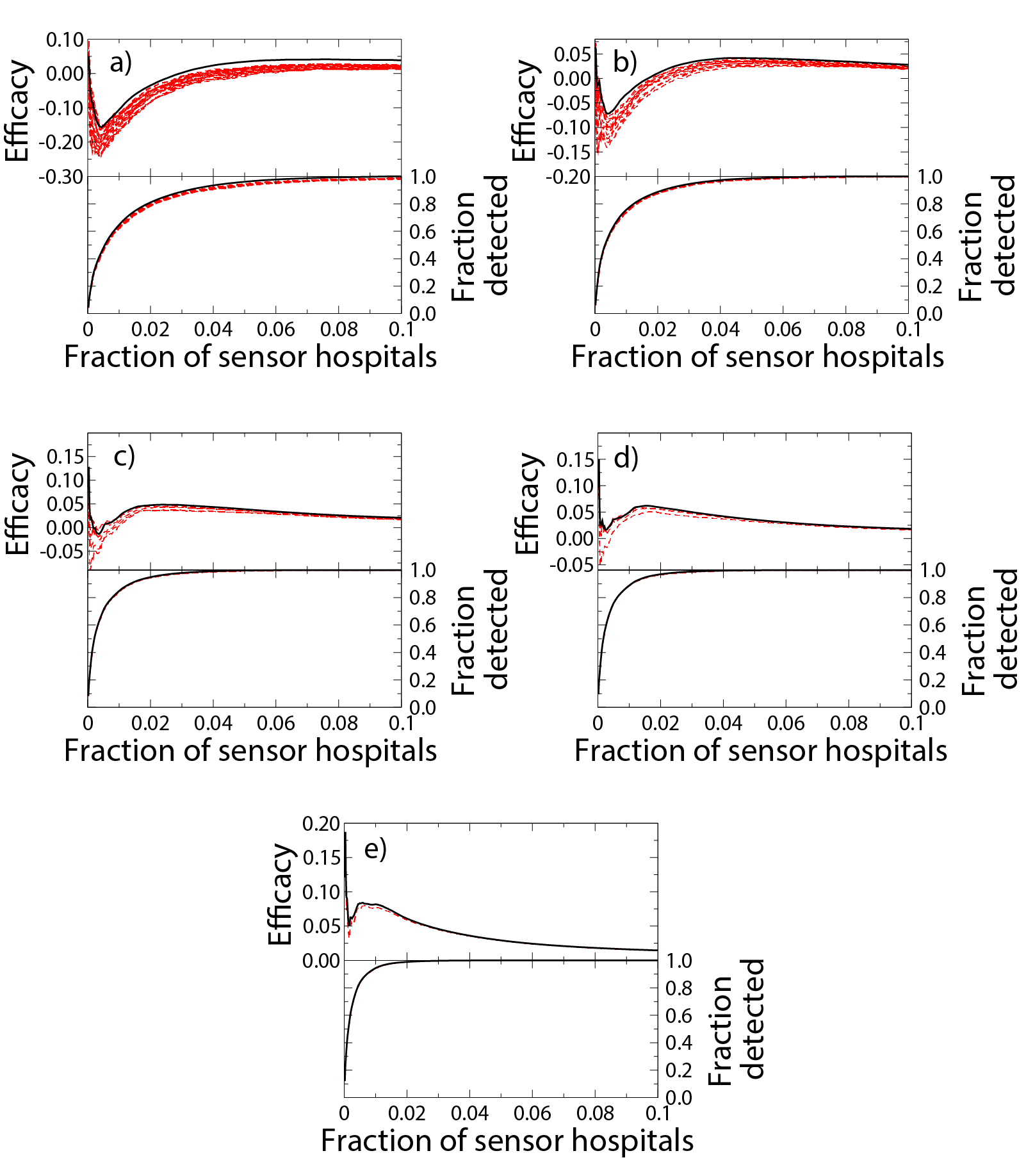}
 \caption{\textbf{ Greedy strategy: information vs. validation sets.} The panels and are arranged as above. Due to the temporal correlation of the data the sensor sets derived from the first slice of data perform comparably to their performance on the training set when applied to the remaining slices of data as test data. Nevertheless when compared to the other strategies this is slightly more variable when compared training and test data results.\label{FigS10}}
 \end{center}
\end{figure}

Although intuitively it seems that the sensor sets would perform worse the greater the temporal separation between the training window and test window, we found that our methods were robust against this separation. Little variation is observed as the validation window gets more and more separated temporally from the training window that was used to construct the sensor sets (see Figs.~\ref{FigS8}-~\ref{FigS10}). This is counterintuitive especially for the sensor set obtained using the greedy algorithm because in principle we are over-fitting our model to the data and consequently this should result in more variability. Nevertheless, temporal correlations in the dynamics of the system make it well behaved in this sense. An important lesson here is that it is possible to determine efficient sensor sets even using outdated data.

\subsection{Effect of the length of the observation period on the sensor set evaluation}

\begin{figure}
\begin{center}
 \includegraphics[width=8.6cm]{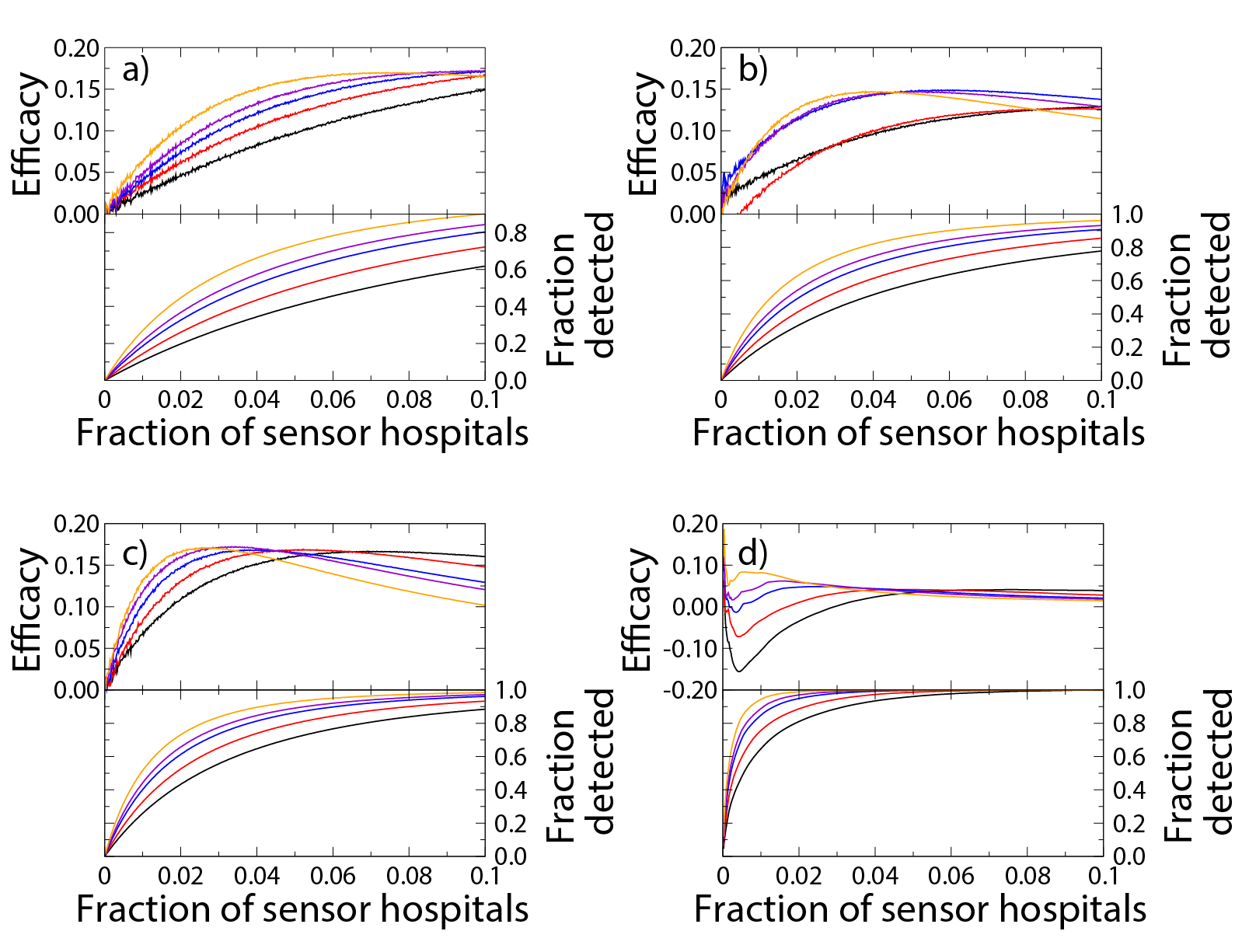}
 \caption{\textbf{ Effect of observation period on the construction of the sensor set.} Efficacy and fraction of detected cases for different lengths of the observation period. \textbf{a)} random strategy, \textbf{b)} out-degree strategy, \textbf{c)} in-degree strategy. \textbf{d)} Greedy strategy, the ``best” sensor set. The different colors correspond to different window widths: 1 month (black), 2 months (red), 4 months (blue), 6 months (purple) and 12 months (orange).
 \label{FigS11}}
 \end{center}
\end{figure}

The validation set approach also enables us to evaluate how the construction of a sensor set is affected by the width of the window used in its construction. From the results in Fig.~\ref{FigS11} it is clear that the wider the window, the smaller the number of sensors needed in order for the sensor set to be optimal. The out-degree strategy is less robust with respect to this metric, and the plots demonstrate a large difference between the curves between 2 and 4 months. The difference is less pronounced between the other curves.

\subsection{List of hospitals included in the sensor sets}

In this section we list the first 26 hospitals included in the in-degree and out-degree strategies, as well as those that arise from the greedy optimization approach.

\vspace{0.2cm}
\textbf{In-degree strategy for the 2-year aggregated network:}
\vspace{0.2cm}
\begin{enumerate}
 \item Saint Marys Hospital, 1216 Second Street SW, Rochester, MN, kin=346, kout=103
 \item Cleveland Clinic Foundation, 9500 Euclid Avenue, Cleveland, OH, kin=286, kout=145
 \item New York-Presbyterian Hospital, 525 East 68th Street, Manhattan, NY, kin=214, kout=118
 \item Mount Sinai Hospital, One Gustave L Levy Place, Manhattan, NY, kin=169, kout=80
 \item St Luke's Episcopal Hospital, 6720 Bertner Avenue, Houston, TX, kin=163, kout=91
 \item Barnes-Jewish Hospital, 1 Barnes-Jewish Hosp Plaza, St. Louis, MO, kin=162, kout=78
 \item Massachusetts General Hospital, 55 Fruit Street, Boston, MA, kin=159, kout=88
 \item Emory University Hospital, 1364 Clifton Road NE, Atlanta, GA, kin=151, kout=71
 \item Methodist Hospital, 6565 Fannin Street, Houston, TX, kin=151, kout=72
 \item University of Alabama Hospital, 619 South 19th Street, Birmingham, AL, kin=147, kout=79
 \item Johns Hopkins Hospital, 600 North Wolfe Street, Baltimore, MD, kin=146, kout=74
 \item UPMC Presbyterian, 200 Lothrop Street, Pittsburgh, PA, kin=146, kout=89
 \item Brigham and Women's Hospital, 75 Francis Street, Boston, MA, kin=142, kout=91
 \item Northwestern Memorial Hospital, 251 East Huron Street, Chicago, IL, kin=141, kout=75
 \item Hospital of the Univ of PA, 3400 Spruce Street, Philadelphia, PA, kin=139, kout=81
 \item Clarian Health Partners, I-65 at 21st Street, Indianapolis, IN, kin=136, kout=81
 \item New York Univ Medical Center, 550 First Avenue, Manhattan, NY, kin=135, kout=46
 \item Kessler Institute for Rehab, 1199 Pleasant Valley Way, Newark, NJ, kin=133, kout=51
 \item Mem Sloan-Kettering Cancer Ctr, 1275 York Avenue, Manhattan, NY, kin=133, kout=52
 \item Duke University Hospital, Erwin Road, Durham, NC, kin=132, kout=67
 \item Rochester Methodist Hospital, 201 West Center Street, Rochester, MN, kin=131, kout=27
 \item Vanderbilt Univ Medical Center, 1211 22nd Avenue South, Nashville, TN, kin=131, kout=77
 \item Baylor Univ Medical Center, 3500 Gaston Avenue, Dallas, TX, kin=131, kout=72
 \item Abbott Northwestern Hospital, 800 East 28th Street, Minneapolis, MN, kin=126, kout=24
 \item Thomas Jefferson Univ Hospital, 111 South 11th Street, Philadelphia, PA, kin=124, kout=75
 \item Lenox Hill Hospital, 100 East 77th Street, Manhattan, NY, kin=123, kout=63
\end{enumerate}

\vspace{0.2cm}
\textbf{Out-degree strategy for the 2-year aggregated network:}
\vspace{0.05cm}
\begin{enumerate}
 \item Cleveland Clinic Foundation, 9500 Euclid Avenue, Cleveland, OH, kout=145, kin=286
 \item New York-Presbyterian Hospital, 525 East 68th Street, Manhattan, NY, kout=118, kin=214
 \item Saint Marys Hospital, 1216 Second Street SW, Rochester, MN, kout=103, kin=346
 \item Brigham and Women's Hospital, 75 Francis Street, Boston, MA, kout=91, kin=142
 \item St Luke's Episcopal Hospital, 6720 Bertner Avenue, Houston, TX, kout=91, kin=163
 \item UPMC Presbyterian, 200 Lothrop Street, Pittsburgh, PA, kout=89, kin=146
 \item Univ of TX M D Anderson Ctr, 1515 Holcombe Boulevard, Houston, TX, kout=89, kin=114
 \item Massachusetts General Hospital, 55 Fruit Street, Boston, MA, kout=88, kin=159
 \item UCSF Medical Center, 500 Parnassus Avenue, San Francisco, CA, kout=81, kin=107
 \item Clarian Health Partners, I-65 at 21st Street, Indianapolis, IN, kout=81, kin=136
 \item Hospital of the Univ of PA, 3400 Spruce Street, Philadelphia, PA, kout=81, kin=139
 \item Mount Sinai Hospital, One Gustave L Levy Place, Manhattan, NY, kout=80, kin=169
 \item University of Alabama Hospital, 619 South 19th Street, Birmingham, AL, kout=79, kin=147
 \item Atlanticare Regional Med Ctr, 1925 Pacific Avenue, Camden, NJ, kout=79, kin=13
 \item Barnes-Jewish Hospital, 1 Barnes-Jewish Hosp Plaza, St. Louis, MO, kout=78, kin=162
 \item Vanderbilt Univ Medical Center, 1211 22nd Avenue South, Nashville, TN, kout=77, kin=131
 \item Florida Hospital, 601 East Rollins Street, Orlando, FL, kout=76, kin=57
 \item Shands at the Univ of Florida, 1600 SW Archer Road, Gainesville, FL, kout=75, kin=106
 \item Northwestern Memorial Hospital, 251 East Huron Street, Chicago, IL, kout=75, kin=141
 \item Thomas Jefferson Univ Hospital, 111 South 11th Street, Philadelphia, PA, kout=75, kin=124
 \item Johns Hopkins Hospital, 600 North Wolfe Street, Baltimore, MD, kout=74, kin=146
 \item Baylor Univ Medical Center, 3500 Gaston Avenue, Dallas, TX, kout=72, kin=131
 \item Methodist Hospital, 6565 Fannin Street, Houston, TX, kout=72, kin=151
 \item Naples Community Hospital, 350 Seventh Street North, Fort Myers, FL, kout=71, kin=27
 \item Emory University Hospital, 1364 Clifton Road NE, Atlanta, GA, kout=71, kin=151
 \item Memorial Hermann Hospital, 6411 Fannin, Houston, TX, kout=71, kin=114
\end{enumerate}

\vspace{0.5cm}

\vspace{0.2cm}
\textbf{Greedy algorithm:}
\vspace{0.25cm}
\begin{enumerate}
 \item Cleveland Clinic Foundation, 9500 Euclid Avenue, Cleveland, OH, kin=286, kout=145
 \item New York-Presbyterian Hospital, 525 East 68th Street, Manhattan, NY, kin=214, kout=118
 \item Saint Marys Hospital, 1216 Second Street SW, Rochester, MN, kin=346, kout=103
 \item Johns Hopkins Hospital, 600 North Wolfe Street, Baltimore, MD, kin=146, kout=74
 \item Massachusetts General Hospital, 55 Fruit Street, Boston, MA, kin=159, kout=88
 \item Univ of TX M D Anderson Ctr, 1515 Holcombe Boulevard, Houston, TX, kin=114, kout=89
 \item Barnes-Jewish Hospital, 1 Barnes-Jewish Hosp Plaza, St. Louis, MO, kin=162, kout=78
 \item Shands at the Univ of Florida, 1600 SW Archer Road, Gainesville, FL, kin=106, kout=75
 \item UCLA Medical Center, 10833 Le Conte Avenue, Los Angeles, CA, kin=116, kout=54
 \item Northwestern Memorial Hospital, 251 East Huron Street, Chicago, IL, kin=141, kout=75
 \item Hospital of the Univ of PA, 3400 Spruce Street, Philadelphia, PA, kin=139, kout=81
 \item Duke University Hospital, Erwin Road, Durham, NC, kin=132, kout=67
 \item Baylor Univ Medical Center, 3500 Gaston Avenue, Dallas, TX, kin=131, kout=72
 \item Emory University Hospital, 1364 Clifton Road NE, Atlanta, GA, kin=151, kout=71
 \item UCSF Medical Center, 500 Parnassus Avenue, San Francisco, CA, kin=107, kout=81
 \item St Joseph's Hosp \& Med Center, 350 West Thomas Road, Phoenix, AZ, kin=58, kout=43
 \item Clarian Health Partners, I-65 at 21st Street, Indianapolis, IN, kin=136, kout=81
 \item Univ of Michigan Hospitals, 1500 East Medical Center Drive, Ann Arbor, MI, kin=113, kout=53
 \item UPMC Presbyterian, 200 Lothrop Street, Pittsburgh, PA, kin=146, kout=89
 \item Vanderbilt Univ Medical Center, 1211 22nd Avenue South, Nashville, TN, kin=131, kout=77
 \item Univ of Washington Medical Ctr, 1959 NE Pacific St, Box 356151, Seattle, WA, kin=74, kout=31
 \item University of Kansas Hospital, 3901 Rainbow Boulevard, Kansas City, MO, kin=95, kout=44
 \item Jackson Memorial Hospital, 1611 NW 12th Avenue, Miami, FL, kin=65, kout=51
 \item OU Medical Center, 1200 Everett Drive, Oklahoma City, OK, kin=69, kout=43
 \item University of Alabama Hospital, 619 South 19th Street, Birmingham, AL, kin=147, kout=79
 \item University of Virginia Med Ctr, Jefferson Park Avenue, Charlottesville, VA, kin=78, kout=48
\end{enumerate}

\end{document}